# Toward Artifact-Free Dermoscopy: Characterization of ISIC 2019 Image Defects for Deep Learning Applications

Victoria Wegley vlwmgz@mst.edu, Joshua Hoog jjh4f4@mst.edu, Tristan Crawford tmch78@mst.edu, Keith Miller kmm734@mst.edu, William Fons wdfgnp@mst.edu, Kacy Pugh kpqhr@mst.edu, Jackson Taylor jhtv8y@health.missouri.edu, Sami Swinfard slsbng@health.missouri.edu, Avigail Fernandes ahftb5@mst.edu, Gehana Patel gnpv5k@missouri.edu, Jason Hagerty jrh55c@mst.edu, William V. Stoecker wvs@umsystem.edu, *R. Joe Stanley stanleyj@mst.edu

*Corresponding author

## Abstract

The International Skin Imaging Collaboration (ISIC) archive offers the largest publicly available collection of dermoscopic images for skin lesion analysis. However, the ISIC 2019 dataset suffers from several acquisition artifacts such as hair occlusions, vignetting, surgical ink markings, and ruler scales, all of which compromise its utility in deep learning workflows by introducing non-biological visual structures. To address these limitations, we present a rigorously annotated version of ISIC 2019 that includes binary masks for key artifact types. Ink marks and vignetting were annotated through systematic manual review and iterative refinement using professional image editing software. For hair and ruler artifacts, we implemented a semi-automatic pipeline based on our previously developed deep learning architecture, ChimeraNet. Automated predictions were then manually corrected as needed for precision. The resulting dataset supports both the development of our own melanoma detection model and broader community research requiring artifact-mitigated dermoscopic images.

## Background & Summary

The ISIC 2019 Challenge dataset comprises 25,331 dermoscopic images spanning nine diagnostic categories: Melanoma, Melanocytic nevus, Basal cell carcinoma, Actinic keratosis, Benign keratosis, Dermatofibroma, Vascular lesion, Squamous cell carcinoma, and None of the others/Out-of-distribution [1]. These images are drawn from multiple international centers and leveraged by researchers worldwide to develop computer-aided diagnostic systems that distinguish lesions requiring expert dermatological review from those that are benign. Despite its scale and impact in training deep learning models, the dataset is impeded by several image-acquisition artifacts, most notably vignetting (dark, circular borders), hair occlusions, surgical ink markings, and ruler scale bars; these artifacts introduce unwanted noise, complicate automated lesion segmentation and classification, and risk models learning spurious correlations rather than true pathological features.

Dermoscopic images in ISIC 2019 exhibit high variability in color saturation and illumination due to differences in dermatoscope light sources, camera settings, and post-processing. Some images suffer from oversaturation where pigment networks and subtle color gradients are clipped, while others are under-exposed or exhibit low contrast, making it difficult to discern fine lesion structures. Vignetting is pervasive throughout the dataset: the circular lens geometry often produces darkened corners or a pronounced black frame that alters the local background tone and can mislead algorithms into using border darkness as a cue. Hair occlusions appear as dark, curvilinear strands crossing over lesions; these hairs obscure critical texture and border information, resulting in fragmented or misleading edge detection in deep learning models. Surgical ink markings, usually applied by clinicians in red, blue, purple, green, or black to outline areas of interest create artifacts by introducing high-contrast, non-biological patterns that can be mistakenly learned by a network as diagnostic features. Figure 1 below presents examples of these artifacts from the ISIC 2019 dataset.

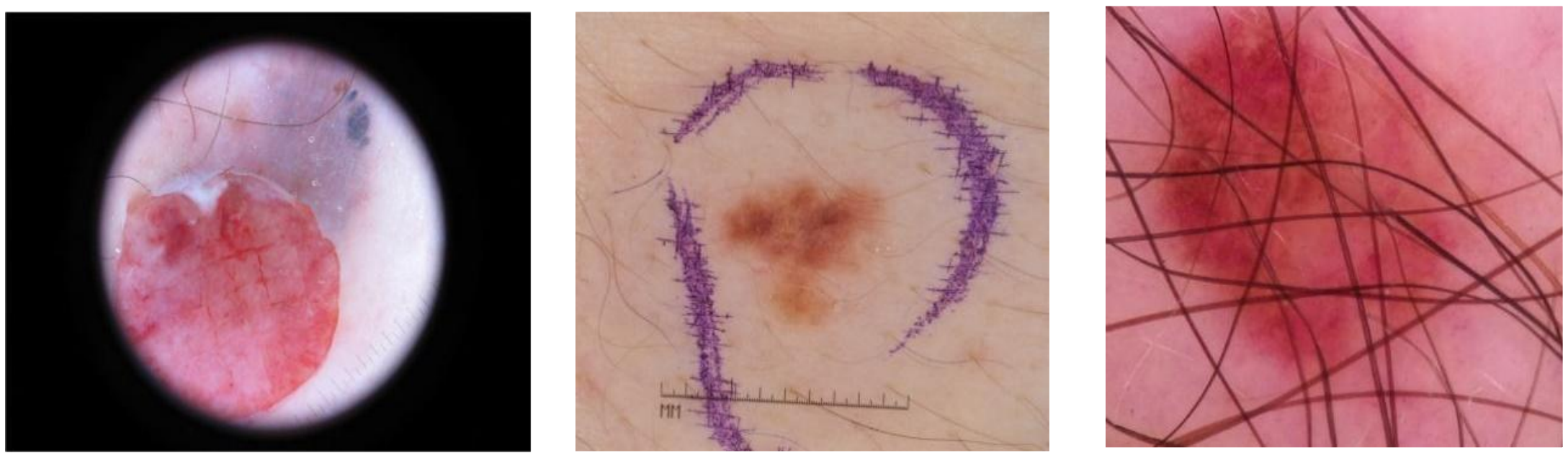

(a) (b) (c)

Figure 1: Examples of artifacts requiring annotations. (a) Hairs. (b) Ink marks. (c) Vignetting.

Similarly, ruler scale artifacts, straight-edged bars, or tick-mark rulers included for size reference inject linear, high-contrast elements into the frame, which models may erroneously associate with particular lesion types. Together, these artifacts along with occasional specular reflections and interference from applied gels degrade the consistency and reliability of the ISIC 2019 images, highlighting the need for careful artifact characterization and removal through annotation and binary masking techniques to prevent interference and memory leaks while training deep learning models.

## Methods

### Binary Masks

Binary masks offer distinct advantages in machine learning applications, particularly by simplifying the problem of the segmentation of artifacts from relevant features into a clear, binary decision for each pixel; an artifact versus background approach is the core of the binary mask approach. Binary representation significantly reduces computational complexity, allowing deep learning models to train faster and more efficiently while also improving model robustness by reducing ambiguity in segmentation tasks, facilitating clearer learning signals during training.

Moreover, the use of binary masks helps deep learning models achieve higher accuracy by providing explicit and unambiguous labels for each pixel. This precision in labeling directly contributes to improved performance in tasks such as semantic segmentation, where clear delineation between classes is crucial. As a result, binary masks enhance model interpretability and enable precise localization of annotated features, leading to more reliable and clinically relevant outputs in dermoscopic image analysis.

**Ink marks**

The ink mark annotations were generated through a systematic review of each image in the dataset to identify instances containing visible ink marks, typically appearing in colors such as black, purple, green, red, or blue. These marks could be located at any position within the image including the surface of the skin, on hair, or present on the lesion itself. Upon identification of images with ink marks, precise masks were manually created to accurately represent the spatial location and shape of these marks.

To produce the ink mark masks, annotators opened each relevant image in an image editing software of their choice, such as Adobe Photoshop, GIMP, or ImageJ, Paint.NET. Initially, a solid black layer was overlaid onto the image, and the opacity was temporarily reduced to enhance the visibility of the underlying original image. Annotators then utilized the brush tool to meticulously outline each ink mark in bright green. The bright green color was specifically selected for its high visibility and contrast against the black background. After completing the outlines, the opacity of the black layer was fully restored, resulting in a binary mask prominently displaying the green outlines of the ink marks with a black background.

Following mask creation, each mask was saved as a TIF file, explicitly labeled to correspond to its original image containing ink marks, and subsequently incorporated into the dataset. This process was consistently applied across all images identified with ink marks, ensuring uniformity and reliability within the dataset. This rigorous approach produced a structured and clearly annotated dataset, aiming to significantly enhance the ability of deep learning models to accurately identify and segment ink marks in dermoscopic images.  Figure 2 shows an original dermoscopy image with ink marks and the associated annotation mask.

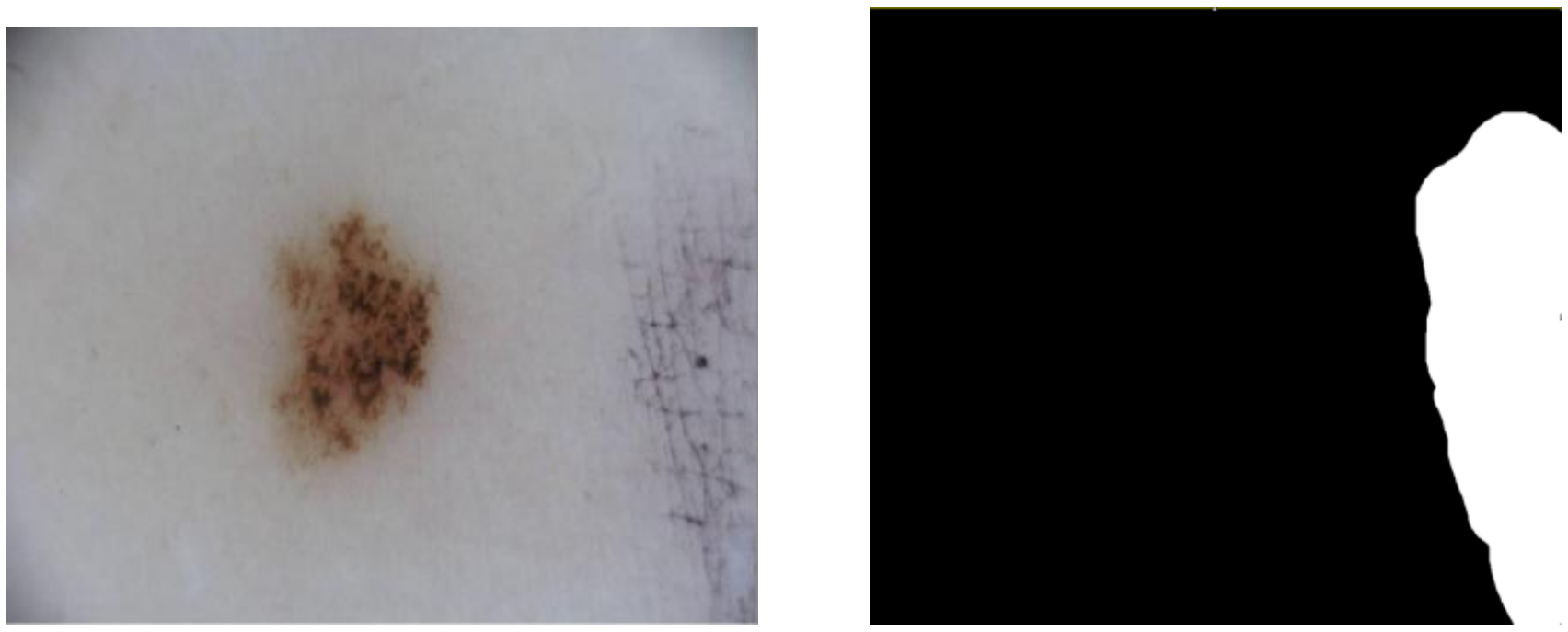

(a) (b)

Figure 2: Ink mark annotation example.  (a) Original skin lesion image. (b) Annotated mask.

**Vignetting**

Vignetting masks were created by identifying images that exhibited characteristic shadows, darkened corners, or circular vignetting throughout the aforementioned systematic review. This common artifact is a result of camera lens limitations, dermatoscope design, and lighting inconsistencies during image acquisition. Annotators first loaded each image identified as containing vignetting into image editing software such as Adobe Photoshop, Paint.NET, GIMP, or Pixlr. Next, a second layer entirely filled with white was added to each loaded image. Annotators then added a small black circle to this layer as a base for manipulation. To facilitate precise adjustments, the opacity of this second white layer was gradually decreased until the original image became visible beneath, and the black circle was just discernible enough for accurate positioning and dimension adjustment.

Annotators then meticulously manipulated the black circle, adjusting its dimensions to fully encompass all non-vignetted regions of the image. The goal was to ensure that the black circle precisely delineated the unaffected central portion, while the white areas accurately represented regions affected by shadows or vignetting. Once adjustments were complete, the opacity of the white layer was restored, resulting in a binary mask clearly distinguishing the vignetted regions in white from non-vignetted regions in black.

Following the completion of mask creation, each mask was saved as a TIF file, clearly labeled as corresponding to the original image identified with vignetting, and added systematically to the dataset. This detailed, iterative procedure was repeated consistently for every vignetted image within the dataset, ensuring the reliability and uniformity of annotations. The rigorously annotated masks are designed to significantly enhance deep learning models' capability to accurately detect, segment, and mitigate vignetting artifacts, thus improving the overall quality and reliability of automated dermoscopic image analysis. Figure 3 presents an example of original skin lesion image with vignetting and the annotated mask obtained through the process described above.

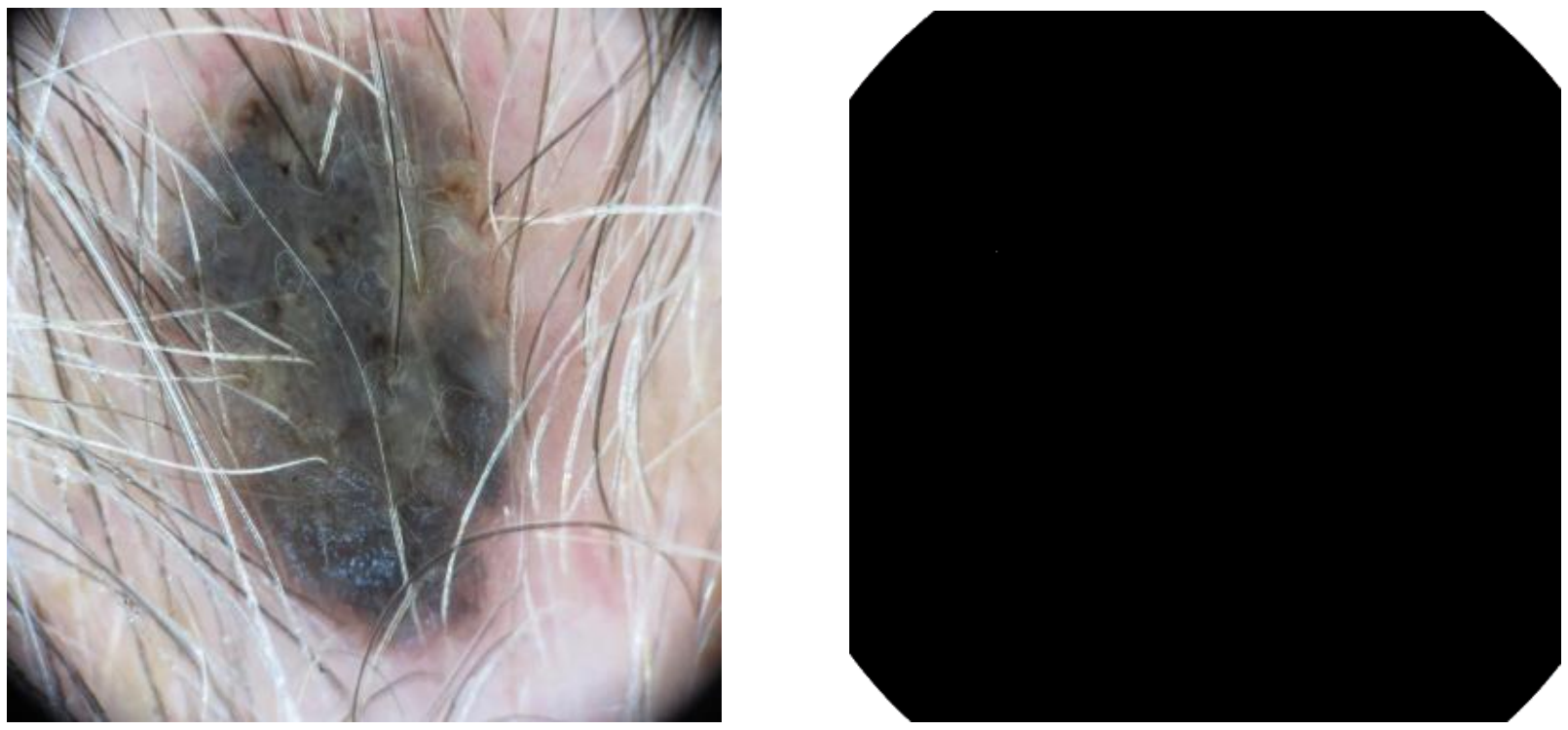

(a) (b)

Figure 3: Vignetting annotation example. (a) Original skin lesion image. (b) Annotated mask.

## Semi-automatic methods

### Hair/ruler marking

Semi-automatic methods were employed to address hair and ruler mark artifacts, which are significant challenges to accurate dermoscopic image segmentation and critical feature identification. Hair artifacts, particularly difficult due to thinness, overlapping strands, faded appearances, or similarity to skin tones and textured lesions, were addressed using a hybrid annotation approach. Initially, automated detection was conducted using ChimeraNet [2], a novel deep learning architecture employing a pre-trained EfficientNet encoder coupled with squeeze-and-excitation residual blocks in its decoder, adapted from the work of Tan and Le [3] and Noorsang et al [4]. The ChimeraNet architecture for hair segmentation uses an encoder-decoder architecture based on a modified EfficientNet model as the encoder network and the decoder network comprised of five squeeze-and-excitation residual blocks. An in-depth presentation of the ChimeraNet architecture and model training for hair and ruler mark segmentation is given in [2]. Annotators can then manually refine the ChimeraNet preliminary segmentation results when necessary, correcting inaccuracies and omissions to ensure precise segmentation across diverse image conditions.

Ruler markings are defined by straight-edged bars or tick marks used for lesion sizing reference, and are addressed through specialized semi-automatic algorithms targeting linear image structures. Following automated detection, meticulous manual inspections were conducted, with annotators making necessary adjustments, particularly in cases involving partially obscured or distorted rulers. This iterative combination of automation and manual intervention streamlined the annotation process, significantly reducing labor while preserving accuracy within the dataset.

Evaluations conducted on the HAM10000 dataset revealed that ChimeraNet achieved superior performance at higher image resolutions (448×448 pixels), attaining an accuracy of 98.23% and a Jaccard index of 0.65, surpassing established deep learning benchmarks such as U-Net and ResUNet-a. Consistent application of the Dice loss function throughout training further optimized performance across metrics [2]. These integrated semi-automatic techniques not only improved annotation efficiency and consistency but substantially enhanced the capability of deep learning models to accurately segment relevant dermoscopic features, significantly minimizing the impact of confounding artifacts. This accuracy paired with the near identical nature of the ISIC 2019 data set positioned ChimeraNet as a key component in efficient and accurate mask development. Figure 4 presents examples of hair and ruler segmentation masks obtained using the ChimeraNet model.

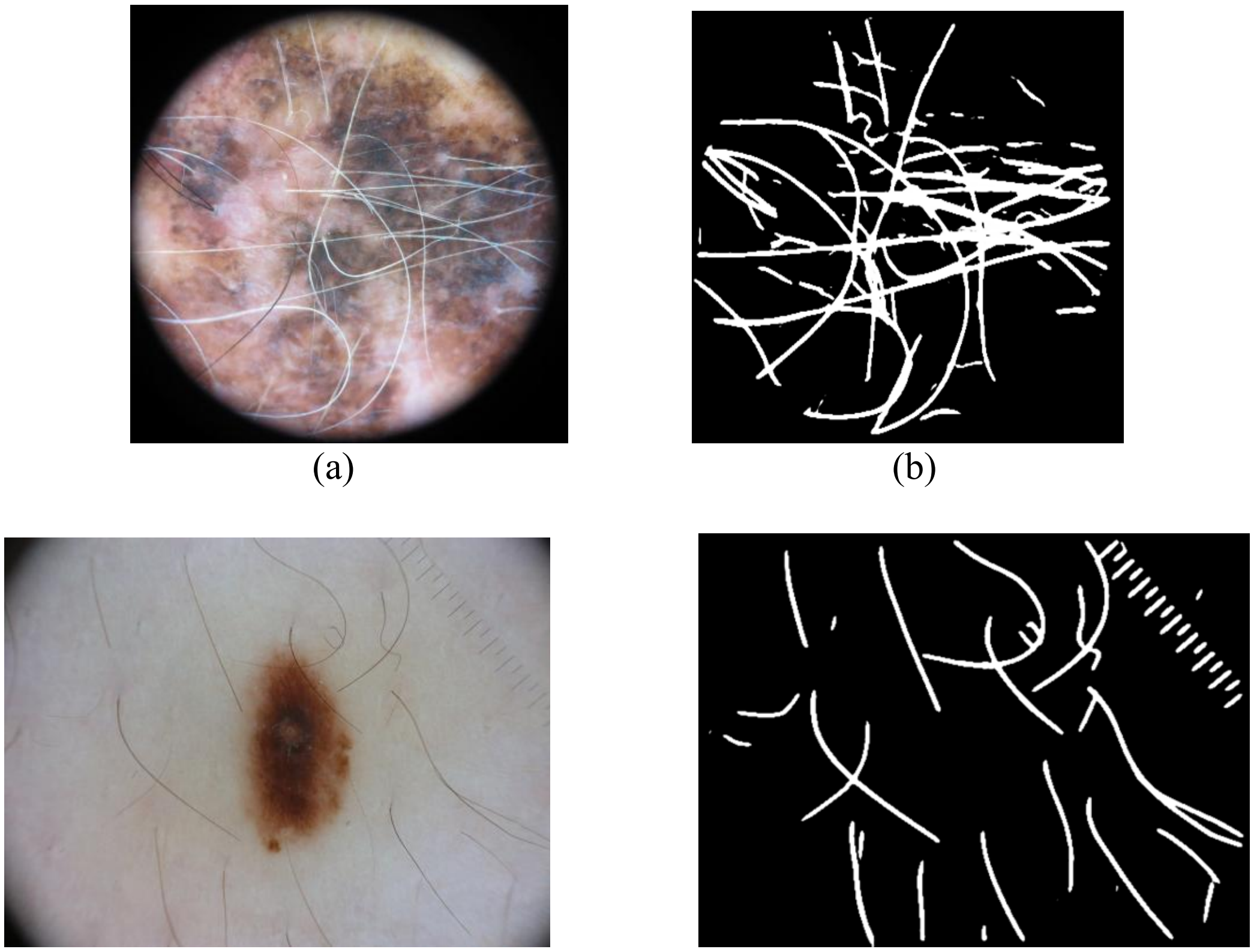
(a) (b) (c) (d)

Figure 4: Hair and ruler segmentation mask examples using ChimeraNet. (a) Original skin lesion image with hairs. (b) Associated hair mask. (c) Original skin lesion image with hairs and ruler mark. (d) Associated hair and ruler mark mask.

## Data Records

The original dermoscopy skin lesion image ISIC 2019 training dataset used in curating the artifact mask sets is at https://challenge.isic-archive.com/data/#2019. Note that ISIC 2019 dataset is a compilation of skin lesion image datasets [5-7]. The mask annotations produced in this study may be accessed at [8].

## Technical Validation

We applied a three-stage quality-control workflow to ensure every artifact mask was anatomically faithful and internally consistent. First, annotators convened regularly to calibrate their interpretations, agreeing on a shared methodology for distinguishing genuine surgical ink from debris and for defining the limits of vignetting shadows. Second, vignetting masks were iteratively refined through empirical, image-by-image visual inspection: masks judged “too tight” (white region intruding into valid background) or “too loose” (insufficient coverage of peripheral darkening) were corrected, while analogous spot-checks removed excessive or missed annotations from ink-mark masks. Finally, the entire repository underwent a full audit; three independent reviewers inspected the ink-mark annotations and two reviewers evaluated the complete vignetting set, resolving discrepancies through joint adjudication with senior investigators – yielding a reproducible, highly accurate annotated data set. From the evaluation process of the ISIC 2019 dataset, 9,187 vignetting, 3,445 ruler mark, and 621 ink mark cases were found with mask images generated. Hair masks and ruler mark masks were automatically generated using the ChimeraNet architecture from Norsang. All original skin lesion and mask images are accessible at the links in the Data Records section above, using the mask naming convention in the section Usage Notes below.

## Context for Data Archive

The goal of this study is to create an archive of specific artifacts present in the 2019 ISIC dermoscopy image dataset that can be independently identified and masked from the skin lesion images to: 1) facilitate improved skin lesion assessment, and 2) promote the recognition of each specific artifact. Yes, these artifacts could be contained in multi-class annotations and could, ultimately, be the format of future annotation masks generated for this archive. Having separate masks for each artifact type does not hinder the development of multi-class AI-based annotation techniques. However, using multi-class annotations within the same mask distracts from the overall goal of expert and algorithm recognition of the individual artifacts for training purposes.

## Usage Notes

The naming convention used to the annotations is:
<ISIC_2019_filename>_<ANNOTATION_TYPE>.tiff

Where ISIC_2019_filename references the original filename in the ISIC 2019 dataset that the annotation corresponds to. The ANNOTATION_TYPE one of the following types:

- mvig (vignetting masks)
- hairmask (hair and ruler masks)
- inkmark (ink mark masks)

Each annotation is saved as a 1-channel binary image with the same image resolution as its corresponding ISIC 2019 data sample. The grayscale value of 255 represents the regions of interest for each annotation, else a value of 0 for all other areas.

## References

[1] “International Skin Imaging Collaboration,” ISIC, https://www.isic-archive.com/ (accessed Nov. 25, 2025).

[2] Lama, N., Kasmi, R., Hagerty, J.R., Stanley, R.J., Young, R., Miinch, J., Nepal, J., Nambisan, A. & Stoecker, W.V. ChimeraNet: U-Net for Hair Detection in Dermoscopic Skin Lesion Images. *J. Digit Imag*. **36(2)**, 526-535 (2023). https://doi.org/10.1007/s10278-022-00740-6. Epub 2022 Nov 16. PMID: 36385676; PMCID: PMC10039207.

[3] Tan, M. & Le, Q. EfficientNet: rethinking model scaling for convolutional neural networks, in International conference on machine learning, 6105–6114 (2019.

[4] Noorsang, K., Sharma, Y. & Kumar, S. Hair and ruler mark detection in dermoscopic images using ChimeraNet: A hybrid encoder-decoder deep learning approach. *Expert Systems with Applications*, **228**, 121395 (2024). https://doi.org/10.1016/j.eswa.2023.121395.

[5] Tschandl P., Rosendahl C. & Kittler H. The HAM10000 dataset, a large collection of multi-source dermatoscopic images of common pigmented skin lesions. Sci. Data 5, 180161 doi.10.1038/sdata.2018.161 (2018)

[6] Noel C. F. Codella, David Gutman, M. Emre Celebi, Brian Helba, Michael A. Marchetti, Stephen W. Dusza, Aadi Kalloo, Konstantinos Liopyris, Nabin Mishra, Harald Kittler, Allan Halpern: "Skin Lesion Analysis Toward Melanoma Detection: A Challenge at the 2017 International Symposium on Biomedical Imaging (ISBI), Hosted by the International Skin Imaging Collaboration (ISIC)", 2017; arXiv:1710.05006.

[7] Hernández-Pérez C, Combalia M, Podlipnik S, Codella NC, Rotemberg V, Halpern AC, Reiter O, Carrera C, Barreiro A, Helba B, Puig S, Vilaplana V, Malvehy J. BCN20000: Dermoscopic lesions in the wild. Scientific Data. 2024 Jun 17;11(1):641.

[8] Dermatology Skin Lesion Image Analysis. Research Data. Scholars’ Mine, Missouri University of Science and Technology. 2026. https://doi.org/10.71674/man1-qa33.

## Author Contributions

The authors were organized into four teams, each specializing in a different annotation task. After training for their assigned annotation type, team members were reallocated as needed to ensure project completion. Throughout every stage, all authors participated in record-keeping and quality control, with final validation carried out by the project managers Dr. Stanley, Dr. Stoecker, and Dr. Hagerty. All authors contributed equally to the writing of this paper.

## Competing Interests

The authors declare no competing interests were a motivation for this work.

## Acknowledgements

This work was supported by grant 1R15CA287304-01from the National Institutes of Health. We thank Dr. V.A. Samaranayake for his role as co-principal investigator, and acknowledge the invaluable contributions of Dr. William V. Stoecker, Dr. Mirna Becevic, and Dr. James Grichnik for their expertise in dataset annotation and clinical guidance. We also gratefully acknowledge Dr. Jason Hagerty and the undergraduate student researchers for their assistance with data collection and analysis and the International Skin Imaging Collaboration for providing access to dermoscopic image datasets.